\documentclass[11pt,secnumarabic,aps,amsmath,amssymb,amsfonts,superscriptaddress,tightenlines,nobibnotes,prd,nofootinbib,twocolumn,dvipsnames]{revtex4}
\usepackage[pass,letterpaper]{geometry}
\usepackage{graphicx,color}
\usepackage{latexsym,amsmath,amsfonts,amssymb,booktabs}
\usepackage[pdfencoding=auto]{hyperref}
\usepackage{url}
\definecolor{MyBlue}{rgb}{0.15,0.15,0.70}
\hypersetup{
colorlinks=true,
citecolor=MyBlue,
linkcolor=MyBlue,
urlcolor=MyBlue
}

\usepackage{bm}
\usepackage[utf8]{inputenc}
\usepackage{subeqnarray}
\usepackage[dvipsnames]{xcolor}
\usepackage[normalem]{ulem}

\newcommand\spart{\;\raise1.0pt\hbox{/}\hskip-6pt\partial}
\newcommand\spartb{\;\overline{\raise1.0pt\hbox{/}\hskip-6pt\partial}}

\newcommand{\be}{\begin{equation}}
\newcommand{\ee}{\end{equation}}
\newcommand{\bea}{\begin{eqnarray}}
\newcommand{\eea}{\end{eqnarray}}
\newcommand{\beal}{\begin{align}}
\newcommand{\eeal}{\end{align}}
\newcommand{\beas}{\begin{subeqnarray}}
\newcommand{\eeas}{\end{subeqnarray}}
\renewcommand\[{\begin{equation}}
\renewcommand\]{\end{equation}}
\newcommand{\dd}{{\rm d}}
\newcommand{\ii}{{\rm i}}

\newcommand{\gr}[1]{\boldsymbol{#1}}

\newcommand{\sourceF}{\mathfrak{u}}
\newcommand{\sourceG}{\mathfrak{v}}

\begin{document}

\title{Optimal Boltzmann hierarchies with non-vanishing spatial curvature}

\author{Cyril Pitrou}
\email[]{pitrou@iap.fr}
\affiliation{Institut d'Astrophysique de Paris, CNRS UMR 7095, 98 bis
  Bd Arago, 75014 Paris, France.}

\author{Thiago S. Pereira}
\email[]{tspereira@uel.br}
\affiliation{Departamento de Física, Universidade Estadual de
  Londrina, Rod. Celso Garcia Cid, Km 380, 86057-970, Londrina,
  Paraná, Brazil.}

\author{Julien Lesgourgues}
\email[]{lesgourg@physik.rwth-aachen.de}
\affiliation{Institute for Theoretical Particle Physics and Cosmology (TTK)\\
RWTH Aachen University, D-52056 Aachen, Germany}

\date{\today}
\begin{abstract}
 Within cosmological perturbation
theory, the cosmic microwave background  anisotropies are usually computed from a Boltzmann hierarchy coupled to the perturbed
Einstein equations. In this setup, one set of multipoles describes the temperature
anisotropies, while two other sets, of electric and magnetic types, describe
the polarization anisotropies. In order to reduce the number of
multipoles types needed for polarization, and thus to speed up the
numerical resolution, an optimal hierarchy has been proposed in the literature for 
Einstein-Boltzmann codes. However, it has been recently shown that the separability between directional and orbital eigenfunctions employed in the optimal hierarchy is not correct in the presence of spatial curvature. We investigate how the assumption of separability affects the optimal hierarchy, and show that 
it introduces relative errors of order $\Omega_K$ with respect to the full hierarchy. Despite of that, we show that the optimal hierarchy still gives 
extremely good results for temperature and polarization angular spectra, with relative errors that are much 
smaller than cosmic variance even for curvatures as large as $|\Omega_K|=0.1$. Still, we find that the
polarization angular spectra from tensor perturbations are significantly altered when using the optimal hierarchy, 
leading to errors that are typically of order $50 |\Omega_K| \%$ on that component.
\end{abstract}

{\small TTK-20-12}

\maketitle

\section*{Introduction}

The radiative transfer of the cosmic microwave background (CMB) is
based on the numerical resolution of a hierarchy of equations coupling
CMB multipoles, together with Einstein equations for the
dynamics of linear metric perturbations. As the CMB is polarized, we have in general a triple hierarchy, with temperature
multipoles (related to intensity $I$), and electric and magnetic type multipoles for linear
polarization (related to $Q$ and $U$ Stokes parameters). In principle,
a fourth hierarchy must be added for circular polarization $V$, but at linear order in perturbation theory it
is not generated by Compton collisions. Hence we have in general as
many hierarchies as Stokes parameters, that is a total
of three coupled hierarchies. An optimal hierarchy valid for flat
Friedmann-Lema\^itre (FL) cosmologies and with only one set
of variables for linear polarization was introduced in \cite{Polnarev1985} and developed further in~\cite{Crittenden:1993ni,Kosowsky:1994cy,Ma1995,Zaldarriaga:1996xe}. 
It was extended to curved FL cosmologies in~\cite{Tram:2013ima}
(TL13 hereafter), leading to a method that was numerically implemented in CLASS\footnote{\url{http://class-code.net}}~\cite{ClassI,ClassII}. The full (i.e., non-optimal) triple hierarchy was 
developed for the flat case in \cite{TAM1} and for the curved case in~\cite{TAM2} and we name it the {\it Total Angular
Momentum} (TAM) hierarchy. Finally, the $1+3$ covariant approach of
\cite{Maartens:1994qq,Gebbie:1999jp,Maartens:1998xg,Challinor:1998xk,Challinor:1999xz,Challinor:2000as,Lewis:2002nc},
which is implemented in CAMB~\cite{Lewis:1999bs,CAMB}, can be mapped to the standard cosmological perturbation
theory~\cite{Bruni:1991kb,Osano:2006ew}. It was found to be equivalent to the TAM approach written in the (matter comoving) synchronous gauge.

Following \cite{TAM1,TAM2}, we summarise in the next section how the triple hierarchy is obtained by expanding temperature and polarization
anisotropies into a complete set of normal modes, valid for any spatial curvature. We then detail in section~\ref{SecOptimal} the key steps 
needed to reduce it to an optimal double hierarchy, following TL13.  Such reduction is based
on a factorization of normal modes into a common orbital function (a
plane wave) and a local angular dependence depending on the normal
mode considered. However, it has been recently shown in \cite{Harmonics} (PP19 hereafter) that for curved
cosmologies, and contrary to what is stated in \cite{TAM2}, this factorization
is not valid. As the optimal hierarchy derivation relies crucially on
this factorization, its implementation in the presence of spatial curvature, as described in TL13, is compromised. Since 
there are hints of mild positive curvature from CMB data~\cite{Aghanim:2018eyx,NatureK}, it becomes crucial to estimate
the errors introduced by the optimal hierarchy in the curved space cases, and this is performed in section~\ref{SecComparison}. We discuss why 
in most cases the error is very small. We describe the modifications implemented in CLASS allowing the user to choose either the TAM or the optimal 
hierarchies when computing angular power spectra. These modifications will be publicly available in a forthcoming CLASS release.

\section{Total angular momentum hierarchy}

\subsection{Normal modes}

Temperature anisotropies depend only the observer's position in spacetime, that is, on the conformal time $\eta$ and the position in
space $\gr{x}$, and on the direction of propagation of the photon $\gr{n}$, which is opposite to the direction of observation. Polarization, which is described by the combinations $Q \pm \ii U$ of Stokes parameters, has the same spacetime dependence.

Temperature and polarization anisotropies are then decomposed along a complete set of normal modes ${}_s M_j^m(\gr{x},\gr{n};\gr{q})$ (with the dependence on the mode $\gr{q}$, the position $\gr{x}$ and the direction of propagation $\gr{n}$ often not written explicitly\footnote{Our normal modes ${}_s M_j^m$ correspond to the ones  of TL13, the  ${}_s \overline{G}^{(jm)}$
  of PP19, and the ${}_s G_j^m$ of \cite{TAM1,TAM2}. The
  mode vector $\gr{q}$ corresponds to $\gr{\nu} \sqrt{|K|}$ in
  PP19, and its norm $q$ is related to the $k$ (used to define tensor harmonics) by $q^2 =
  k^2+(1+|m|)K$.}), which are projections of tensor valued harmonics, as
\be\label{ThetaG}
\Theta = \sum_{jm}\int \frac{\dd^3 \gr{q}}{(2\pi)^3}\,\Theta_j^m(\gr{q},\eta) \,{}_0 M_j^m(\gr{x},\gr{n};\gr{q})\,,
\ee
and
\begin{align}\label{EBG}
&Q\pm \ii U = \sum_{jm}\int \frac{\dd^3 \gr{q}}{(2\pi)^3}\,\\
&\qquad \times \left[E_j^m(\gr{q},\eta) \pm \ii   B_j^m(\gr{q},\eta)
\right]\,{}_{\pm 2} M_j^m(\gr{x},\gr{n};\gr{q})\,.\nonumber
\end{align}
Here $m\in[-2,2]$ is the mode index, standing respectively for scalars ($m=0$), vectors ($|m|=1$) and tensors ($|m|=2$), while $j\geq0$ is the multipole index.
The normal modes depend on curvature $K$ of spatial sections\footnote{Recall that $|K|=\ell_c^{-2}$, where $\ell_c$ is the curvature length of spatial sections.}, and are expressed in terms of radial functions and spin-weighted spherical harmonics. A comprehensive set of their properties is collected in PP19.

\subsection{Hierarchy}

The evolution of anisotropies is governed by the Boltzmann equation 
\beas\label{BoltzmannTheta}
\left(\partial_\eta + \gr{n} \cdot {\bm \nabla} + \tau'\right)\Theta &=& {\cal C}_\Theta + {\cal G}\,,\\
\left(\partial_\eta+ \gr{n} \cdot {\bm \nabla}+ \tau'\right) (Q\pm\ii U) &=& {\cal C}_{Q\pm\ii U}\,,
\label{BoltzmannQU}
\eeas
where $\tau'$ is the Compton scattering rate. The function ${\cal G}$ accounts for the gravitational effects due to metric
perturbations, and it is decomposed on normal modes similarly to
\eqref{ThetaG}, hence defining the multipoles ${\cal G}_j^m$. The only
non-vanishing gravitational sources satisfy $j\leq2$ (with $|m|\leq j$), and can be found in e.g.~\cite{TAM2,Tram:2013ima,Harmonics}.

The collision terms ${\cal C}_\Theta$ and $ {\cal C}_{Q\pm\ii U}$ are also expanded on normal modes, similarly to \eqref{ThetaG} and
\eqref{EBG}, hence defining the multipoles $
{}^\Theta{\cal C}_j^m $, ${}^E{\cal C}_j^m $ and ${}^B{\cal C}_j^m
$. The only non-vanishing contributions are also restricted to $j
\leq 2$, and can be found in \cite{TAM2}.

Using
\bea\label{MasterGstreaming}
&&-\gr{n}\cdot  {\bm \nabla} \left({}_s M_j^m \right)  = \frac{\ii q m s}{j(j+1)}  \,{}_s M_j^m \\
&&+\frac{1}{2j+1}\left[-{}_s \kappa_j^m\,  {}_s M_{j-1}^m +{}_{s} \kappa_{j+1}^m  \,{}_s M_{j+1}^m \right]\nonumber
\eea
with coupling coefficients
\be\label{Defkappaslm}
{}_s \kappa_j^m \equiv
\sqrt{\frac{(j^2-m^2)(j^2-s^2)}{j^2}}\sqrt{q^2-K j^2}\,,
\ee
we obtain immediately the TAM hierarchy~\cite{TAM2}
\begin{align}\label{Hierarchy}
&\partial_\eta \Theta_j^{m} = {\cal G}_j^m +{}^\Theta {\cal C}_j^m   -\tau' \Theta_j^m \\
  &\qquad \qquad +\left[\frac{{}_0 \kappa_j^m}{2j-1} \Theta_{j-1}^m - \frac{{}_0 \kappa_{j+1}^m}{2j+3} \Theta_{j+1}^m \right], \nonumber\\
&\partial_\eta E_j^{m} = {}^E {\cal  C}_j^m -\tau' E_j^m\nonumber\\
 &+\left[\frac{{}_2 \kappa_j^m}{2j-1}  E_{j-1}^m - \frac{{}_2 \kappa_{j+1}^m}{2j+3}
  E_{j+1}^m -\frac{2m q}{j(j+1)}   B_j^m\right],\nonumber\\
  &\partial_\eta B_j^{m} = {}^B{\cal C}_j^m -\tau' B_j^m\nonumber\\
   &+ \left[\frac{{}_2 \kappa_j^m}{2j-1} B_{j-1}^m - \frac{{}_2 \kappa_{j+1}^m}{2j+3}B_{j+1}^m +\frac{2m q}{j(j+1)}  E_j^m\right]. \nonumber
\end{align}

A Boltzmann code must solve this set of equations, along with the evolution of metric
perturbations (in a given gauge) which enter the gravitational
sources, for various values of the mode magnitude $q$. The temperature and polarization
angular spectra are then obtained from convolutions with the initial
perturbations power spectra, and take simple forms for statistically isotropic initial conditions, see e.g. section 
2.E of \cite{TAM2} or section 3.4 of \cite{Lesgourgues:2013bra}.
In these equations, $m\in[-j,j]$ can be positive or negative, but since hierachies with the same $j$ and opposite $m$ return identical results, 
calculations can be performed for $m\geq0$ only.

\subsection{Integral solutions}

Since the Boltzmann hierarchy is infinite in $j$, we must in practice
truncate at a $j_{\rm max}$ sensibly larger than the maximum $j$ we
are interested in, so as to avoid errors introduced by the truncation.
It is in practice much faster to solve only for a limited number of
multipoles, that is to truncate at a low $j_{\rm max}$, and to reformulate the solutions of the Boltzmann hierarchy
 as an integral on sources involving these lowest multipoles. This line of sight method was first introduced in \cite{TAM1,Seljak:1996is}. It is indeed
 found that the solutions of the hierarchy \eqref{Hierarchy} are 
\begin{align}\label{IntSol}
\frac{\Theta_j^m(\gr{q},\eta_0)}{2j+1} &=  \int_0^{\eta_0} \dd \eta {\rm e}^{-\tau} \!\!\!\!\! \sum_{j'=m, ..., 2}  \! {}_0 \epsilon_j^{(j'm)}(\chi;q)\\
&\qquad \times \left[{}^\Theta{\cal C}_{j'}^m(\gr{q},\eta)  + {\cal G}_{j'}^m(\gr{q},\eta)  \right]\,,\nonumber\\
\frac{ E_j^m(\gr{q},\eta_0)}{2j+1} &=  \int_0^{\eta_0} \dd \eta {\rm  e}^{-\tau} \,  {}_2 \epsilon_j^{(2,m)}(\chi;q) \, {}^E{\cal  C}_{2}^m(\gr{q},\eta)\,\,,\nonumber\\
\frac{B_j^m(\gr{q},\eta_0)}{2j+1} &=  \int_0^{\eta_0} \dd \eta {\rm e}^{-\tau} \,{}_2 \beta_j^{(2,m)}(\chi;q) {}^E{\cal C}_{2}^m(\gr{q},\eta)\,,\nonumber
\end{align}
where $\chi = \eta_0-\eta$ is the radial distance. The optical depth is $\tau$ (such that $\dd
\tau/\dd \eta = -\tau'$ and with $\tau(\eta_0)=0$) and the ${}_s \epsilon_\ell^{(jm)}(\chi;q)$ and ${}_s \beta_\ell^{(jm)}(\chi;q)$
are the electric and magnetic type radial functions (reported in section 4 of PP19), initially introduced in \cite{Tomita1982,Abbott1986,TAM2} for curved spaces\footnote{${}_0 \epsilon_\ell^{(jm)}$ corresponds to $\phi_\ell^{(jm)}$ in \cite{TAM2,Tram:2013ima},
and ${}_2 \epsilon_\ell^{(2,m)}, {}_2 \beta_\ell^{(2,m)}$ to
$\epsilon_\ell^{(m)},\beta_\ell^{(m)}$. Note also that ${}_s \alpha_\ell^{(jm)}$, ${}_s \epsilon_\ell^{(jm)}$ and ${}_s \beta_\ell^{(jm)}$ correspond to ${}_s \bar\alpha_\ell^{(jm)}$, ${}_s \bar\epsilon_\ell^{(jm)}$ and ${}_s \bar\beta_\ell^{(jm)}$ of PP19.}. 
These results follow from the structure of the Boltzmann
equation~\eqref{BoltzmannTheta}, once written in an integral form (for
instance, for temperature, $ \dd/\dd \tau ({\rm e}^{-\tau}\Theta) = {\rm e}^{-\tau}[ {\cal C}_\Theta + {\cal G}]$), and
using the Rayleigh expansion (e.g. Eq.~(7.39) of PP19) to
express the normal modes of gravitational and collisional sources in
terms of the normal modes evaluated at the observer (that is at
$\chi=0$), see section 7.4 of PP19 for more details.

Finally, the unlensed angular spectra  $C_\ell^{X}$ for $X\in[TT,TE,EE,BB]$ are given by the integral over $\gr{q}$ of products of $\Theta_j^m(\gr{q},\eta_0)$, $E_j^m(\gr{q},\eta_0)$, $B_j^m(\gr{q},\eta_0)$ multiplied by the primordial power spectra.

\subsection{Hierarchy truncation}

The radial functions involved in the integral solutions~\eqref{IntSol} have
a variety of recursive properties. In particular, setting $s=\ell$ in Eq. (D.5) of PP19, and promoting the changes $s\leftrightarrow m$ and $j\leftrightarrow \ell$ by means of Eqs.(3.26) and (3.27) of the same reference, we can show that (see also section
5.4.5 of \cite{RiazueloPhD})
\bea\label{MagicalTruncation}
&&\left(\frac{\dd}{\dd \chi} +(\ell+1+m) \cot_K(\chi)\right)
\,{}_{s}\alpha^{(j=m,\pm m)}_\ell \\
&&-\frac{{}_{s}\kappa_{\ell}^m}{(\ell-m)} \, {}_{s}
\alpha^{(j=m,\pm m)}_{\ell-1} \pm  \ii \frac{s \nu}{\ell}
{}_{s}\alpha^{(j=m,\pm m)}_\ell=0\,,\nonumber
\eea
where ${}_{\pm s}\alpha^{(j,m)}_\ell = {}_s\epsilon^{(j,m)}_\ell\pm\ii\, {}_s\beta^{(j,m)}_\ell$ and 
$\cot_K(\chi)$ corresponds to either $\sqrt{|K|}\coth (\chi \sqrt{|K|})$, $\sqrt{K}\cot (\chi \sqrt{K})$, or $1/\chi$ when $K$ is smaller than, greater than
or equal to zero, respectively. One can deduce from \eqref{IntSol} and (\ref{MagicalTruncation}) that
\begin{enumerate}
\item if non-vanishing sources are located only very deep in the past (at distances such that $\chi = \eta_0-\eta \simeq
  \eta_0$),
\item if we can ignore sources with $j>|m|$ (which is in general not the case),
\end{enumerate}
then the temperature multipoles satisfy 
\bea\label{Truncation1}
\partial_\eta \Theta^{m}_j&\simeq& -(j+1+m)\cot_K(\eta) \Theta^{m}_j\\
&+&\frac{{}_{0}\kappa_{j}^m}{(j-m)} \,\frac{2j+1}{2j-1}
\Theta^{m}_{j-1} \,.\nonumber
\eea
Similarly, and using the fact that ${}_s \alpha_\ell^{(j,m)} =
{}_m\alpha_\ell^{(j,s)}$ in \eqref{MagicalTruncation}, we find 
under the same first assumption (but relaxing the second one) that the polarisation multipoles satisfy
\bea\label{Truncation2}
\partial_\eta E_j^m &\simeq& -(j+3)\cot_K(\eta) \,E^{m}_j \\
&+&\quad\frac{{}_{2}\kappa_{j}^m}{(j-2)} \, \frac{2j+1}{2j-1}
E^{m}_{j-1} + \frac{m q}{j}   B^{m}_j\,,\nonumber
\eea
with $B_j^m$ satisfying the same approximate relation with replacements $E_j^m \to B_j^m$ and $B_j^m \to - E_j^m$. 

Equations \eqref{Truncation1} and \eqref{Truncation2} are only approximate, but they can be used in practice to truncate the hierarchy at a $j_{\rm max}$, 
so as to minimize spectrum reflection that a direct truncation of
\eqref{Hierarchy} would induce.

\section{Optimal hierarchy}\label{SecOptimal}

It has been conjectured in \cite{TAM2} and assumed in \cite{Tram:2013ima} that the normal modes can be separated into the product of a local angular structure 
and some eigenmode functions $\Delta$ normalized to $|\Delta|=1$:
\be\label{WrongFactor}
{}_s M_j^m = (-\ii)^j\sqrt{\frac{4\pi}{2j+1}}\, {}_s Y_{j}^m(\gr{n})
\Delta(\gr{x}, \gr{q})~.
\ee
In reality, this property is lost in the presence of spatial curvature, as detailed in section 6.7 of PP19. In the flat case, where the function $\Delta=\exp(\ii \gr{q}\cdot\gr{x})$ consists of ordinary plane waves, a series of simplifications leads to the optimal hierarchy, which we now review.

First, for temperature, one can expand the non scalar perturbations ($m \neq 0$) using the same normal modes as for scalar perturbations
($m=0$). Then, instead of using the $\Theta_j^m$, one can use a new set of multipoles $F_j^m$ defined  by
\be\label{DefineFjmHierarchy}
\sum_j \Theta_j^m \,{}_0M_j^m \propto Y_m^m\sum_j (2j+1)F_j^m \,
{}_0M_j^0\, \,.
\ee
Note that the factor $(2j+1)$ and the global pre-factor (not shown here) are pure conventions in the definition of $F_j^m$, and that the new multipoles are defined for $j \geq 0$ (unlike $\Theta_j^m$ which is defined for $j\geq |m|$). If the factorization of eq.~(\ref{WrongFactor}) holds, this relation is unchanged when replacing ${}_{0} M_j^m \to {}_{0} Y_j^m$. Then, using the orthogonality relation of spherical harmonics, one finds that the $\Theta_j^m$ can be
deduced from the $F_j^m$ using Gaunt integrals (angular integrals over
three spin-weighted spherical harmonics). These relations are
collected in appendix B of TL13. For scalar modes, since $Y^0_0=1/\sqrt{4\pi}$, the Gaunt integral becomes trivial, such that the mutipoles $F_j^0$ and $\Theta_j^0$ are just related by numerical factors. We see that for temperature, switching to the optimal hierarchy amounts in expanding along the basis of angular functions $Y_m^m Y_j^0$ instead of $Y_j^m$. This explains why the source terms remain compact: given the contraction rules of spherical harmonics, the source terms in the two hierarchies are simply related through Clebsh-Gordan coefficients. For instance, for the gravitational source terms ${\cal G}_j^m$, restricted to $|m|\leq j \leq 2$, we immediately see that ${\cal G}_m^m$ sources $F_0^m$ (since they are both factors of $Y_m^m \propto Y_m^m Y_0^0$),  that ${\cal G}_j^0$ sources $F_j^0$ (both factors of $Y_j^0 \propto Y_0^0 Y_j^0$), and finally that ${\cal G}_2^{\pm1}$  sources $F_1^{\pm 1}$ (since $Y_2^{\pm1} \propto Y_1^{\pm1} Y_1^0$).
The source coming from Thomson scattering also remains simple because the baryon velocity has a dipolar structure, $j=1$, 
that can only source $F_0^{\pm1}$ and $F_1^0$ (following the same reasoning as for ${\cal G}_1^m$).

Second, for polarization, the problem can be simplified by use of symmetries. The Stoke parameter combinations ${Q+\ii U}$ and ${Q-\ii U}$ both start from vanishing initial conditions and grow according to the Boltzmann equations (\ref{BoltzmannQU}), which differ only at the level of the collision terms ${\cal C}_{Q+\ii U}$ and ${\cal C}_{Q-\ii U}$. However the Thomson scattering cross section has a quadrupolar structure giving
\be\label{CollQiU}
{\cal C}_{Q\pm\ii U} =-\sqrt{6}\tau' \!\! \sum_{m=-2}^2 \int \!\! \frac{\dd^3 \gr{q}}{(2\pi)^3}\, P^{(m)} {}_{\pm 2} M_2^m\,,
\ee
with $P^{(m)}(\gr{q},\eta)  \equiv \left(\Theta_2^m-\sqrt{6}\,E_2^m\right)/10$. In general, these source terms give no useful relation between ${Q+\ii U}$ and ${Q-\ii U}$. However, if the
factorization property \eqref{WrongFactor} holds, ${\cal C}_{Q\pm\ii U}$ can be written as 
\be\label{CollQiUopt}
{\cal C}_{Q\pm\ii U}  \propto \sum_{m=-2}^2 \left[  \int \frac{\dd^3 \gr{q}}{(2\pi)^3}\, P^{(m)} \Delta \right] {}_{\pm 2} Y_2^m ~,
\ee
where the bracketed integral only depends on $(\gr{x}, \eta)$ and is the same for $Q+iU$ and $Q-iU$. 
Thus each mode $m$ sources identical contributions to ${Q+\ii U}$ and ${Q-\ii U}$ up to a ratio  ${}_{- 2} Y_2^m/{}_{+ 2} Y_2^m$ that only depends on the direction $\gr{n}$. By taking the sum and the difference of eqs.~(\ref{BoltzmannQU}), one reaches similar conclusions for $Q$ and $\ii U$: each mode $m$ sources identical contributions to the Stokes parameters up to a factor
\be\label{EBratio}
\frac{{\ii U}}{Q}=\frac{{}_2Y_2^m - {}_{-2}Y_2^m }{{}_2Y_2^m + {}_{-2}Y_2^m}~.
\ee
Note that (\ref{EBratio}) also holds for scalar modes, for which ${}_2Y_j^0 \!\! = \!\! {}_{-2}Y_j^0$ and $U$ is not sourced. When computing CMB spectra, we consider statistically independent initial conditions for each mode $m$, and thus solve the Boltzmann equations for one mode $m$ at a time. Thus we can solve only for $Q$ and assume  that $\ii U$
is given by eq.~(\ref{EBratio}).

In general, the sum of the two equations \eqref{EBG} shows that $Q$ is
related to polarization electric and magnetic multipoles as
\be
Q = \frac{1}{2}\sum_{jm} \int \frac{\dd^3 \gr{q}}{(2\pi)^3} \left(E_j^m
{\cal E}_j^m + \ii B_j^m {\cal B}_j^m \right)\,,
\ee
where we have defined the E and B type normal modes
\bea
{\cal E}_j^m &\equiv& \left({}_2M_j^m + {}_{-2}M_j^m\right)\,,\\
{\cal B}_j^m &\equiv&  \left({}_2M_j^m - {}_{-2}M_j^m\right)\,.
\eea
In the optimal scheme, $Q$ can instead be expanded in a single hierarchy of multipoles $G_j^m$ that involves the same normal modes ${}_0M_j^0$ as the temperature expansion:
\bea\label{DefineGjmHierarchy}
\sum_{j} \! E_j^m {\cal E}_j^m \!\!+\! \ii B_j^m {\cal B}_j^m \!
\propto  \! \tilde{\cal E}^m \!\! \sum_j \! (2j\!\!+\!\!1) G_j^m  {}_0M_j^0\,,
\eea
where $\tilde{\cal E}^m(\gr{n})$ is chosen to simplify the Boltzmann hierarchy as much as possible. Again, if the factorization property \eqref{WrongFactor} holds, this relation can be written with ${}_{\pm 2} M_j^m \to {}_{\pm 2} Y_j^m$, and if $\tilde{\cal E}^m$ is a spherical harmonic, we can find the relation between $G_j^m$ and $(E_j^m, B_j^m)$ using Gaunt integrals, as detailed in appendix B of TL13. According to eq.~(\ref{CollQiU}), ${\cal C}_Q(\gr{n}) \propto  {\cal E}_2^m(\gr{n})$. Thus, for $m\neq0$, choosing $\tilde{\cal E}^m \propto {\cal E}_2^m$ leads to a simple Boltzmann hierarchy. Indeed, in the right-hand side of eq.~(\ref{DefineGjmHierarchy}), scattering can only source the multipoles such that $Y_j^0$ is direction-independent, that is, $G_0^m$. For $m=0$, in order to recover the equations reported in \cite{Ma1995}, TL13 chose $\tilde{{\cal E}}^0$ to be a constant factor (instead of ${\cal E}_2^0$) such that the multipoles $G_j^0$ relate to $Q$ exactly as $F_j^0$ relate to $\Theta$. This choice comes however at the expense of an additional source term for $G_2^0$ in the hierarchy, and of less straightforward relations between $E_j^0$ and the $G_j^0$.

Having reduced the expansion on the simpler normal modes ${}_0 M_j^0$, one gets temperature and polarization hierarchies that are both very similar to the scalar temperature hierarchies of the TAM method,
\beas
\partial_\eta F_j^m &=& \frac{1}{2j+1}\left({}_0 \kappa_j^0
                      F_{j-1}^m- {}_0 \kappa_{j+1}^0 F_{j+1}^m\right)\nonumber\\
  &&\quad -\tau' F_j^m + \sourceF_j^m\,,\\
\partial_\eta G_j^m &=& \frac{1}{2 j+1}\left({}_0 \kappa_j^0
                      G_{j-1}^m- {}_0 \kappa_{j+1}^0 G_{j+1}^m\right)\nonumber\\
  &&\quad -\tau' G_j^m + \sourceG_j^m\,,
\eeas
with the sources $\sourceF_j^m,\sourceG_j^m$ and exact definitions for the $F_j^m,G_j^m$ given in TL13.
Also, since the free-streaming part has been reduced in all cases to the same form as scalar temperature multipoles, the hierarchies
for  $F_j^m,G_j^m$ are truncated using \eqref{Truncation1} with $m=0$ in all cases, that is Eq. (2.34) of TL13.

Finally, the temperature and polarization spectra can be computed using eq.~\eqref{IntSol}, with the same radial functions as in the TAM method, but with the expression of the source functions $^{\Theta}{\cal C}_j^m$, $^{E}{\cal C}_j^m$ and ${\cal G}_j^m$ derived in the optimal hierarchy.

The optimal hierachy equations were already derived in TL13, but the goal of this section was to show explicitly that, at various steps in the derivation, it is necessary to assume the factorization ansatz of eq.~\eqref{WrongFactor}. As found in PP19, in the presence of spatial curvature, this factorization does not hold, such that the optimal hierarchy should not be used in principle.

\section{Comparison of hierarchies.}\label{SecComparison}

\subsection{Implementation in CLASS}

Previous versions of the CLASS code were only using the optimal hierarchy. For the purpose of comparing the two schemes, we have implemented both of them, with a new input parameter {\tt hierarchies = optimal, tam}. Our modifications will be available in the next release of the code (v3.0). For the first three multipoles of the scalar temperature hierarchy, instead of following ($\Theta_0^0$, $\Theta_1^0$, $\Theta_2^0$) or ($F_0^0$, $F_1^0$, $F_2^0$), the code follows three components of the  perturbed photon stress-energy tensor that match the conventions of \cite{Ma1995}:
\beas
\delta_\gamma &=& F_0^0 = 4 \Theta_0^0~, \\
\theta_\gamma &=& \frac{3k}{4} F_1^0 = k \Theta_1^0~,  \\
\sigma_\gamma &=& \frac{1}{2s_2} F_2^0 =  \frac{2}{5s_2} \Theta_2^0~,  
\eeas
with $k=\sqrt{q^2-K}$ and $s_2 = \sqrt{1-3K/k^2}$. 
For all other multipoles and modes, the code follows the quantities ($F_\ell^m$, $G_\ell^m$) in the optimal mode and ($\Theta_\ell^m$, $E_\ell^m$, $B_\ell^m$) in the TAM mode.

We have discussed the two hierachies in the context of photon anisotropies, but the same formalism applies to decoupled massless or massive neutrinos, or more generally to ultra-relativistic species ({\it ur}) and non-cold dark matter ({\it ncdm}), as they are called in CLASS). The only difference in such cases is the absence of both polarization and collision terms. 

For scalar modes, in absence of polarization, the TAM and optimal hierarchies are mathematically equivalent, even when $K\neq0$. This can be seen in the definition of the $F_j^0$ multipoles in equation (\ref{DefineFjmHierarchy}). With $m=0$, given that $Y_0^0=1/\sqrt{4 \pi}$, we see that the expansions in $\Theta_j^0$ and in $F_j^0$ are performed along the same normal modes $_{0}M_j^0$. Then, even if $_{0}M_j^0$ is not separable in curved space, the optimal hierarchy can be obtained from the TAM one by replacing 
$\Theta_j^0 \to (2j+1) F_j^0$ (up to a constant factor $\frac{1}{4}$ coming from an arbitrary choice of normalization in \eqref{DefineFjmHierarchy}). For photons, there is still a difference in the temperature evolution, coming from the fact that the temperature hierarchy couples to distinct polarisation hierarchy(ies). But this is not the case for the {\it ur} and {\it ncdm} species, and thus there  is no need to implement explicitly the TAM hierarchy for them.

On the other hand, for tensor modes, we expect the optimal hierarchy to be only approximate in the curved case, due to the non-separability of the normal modes $_{0}M_j^2$, which implies that $\Theta_j^2$ and $(2j+1) F_j^2$ are not exactly related by Gaunt integrals. This is potentially relevant for the calculation of the spectra of CMB anisotropies, since photon and neutrino are coupled gravitationally through their shear tensors.

In both CLASS and CAMB, for tensor modes, the impact of massive neutrinos (or more generally {\it ncdm}) perturbations on the CMB angular spectra can be accounted in two ways: (i) either using the full Boltzmann hierarchy of {\it ncdm} perturbations discretized on a grid in momentum space; or (ii) by splitting {\it ncdm} at each time $\eta$ in two components: an ultra-relativistic component with density $\rho=3p_{ncdm}$, treated as an enhancement of the {\it ur} species and thus coupled gravitationally to the photons, and a non-relativistic component with density $\rho=\rho_{ncdm} - 3p_{ncdm}$, assumed to have a negligible shear and thus no gravitational coupling with photon tensor perturbations. The second scheme is faster and accurate enough (at least for neutrinos becoming non-relativistic after photon decoupling)  for being the default in CLASS. In that case, for tensor modes, the code follows the {\it ur} perturbations but not the {\it ncdm} ones.

Here, we limit our analysis to the case where this approximation is used. Thus, we coded the two hierarchies for {\it ur} tensor perturbations, but not for the {\it ncdm} tensor perturbations. Depending on the used scheme, the code follows either the multipoles $F_{ur,j}^{~~2}$ (optimal) or $\Theta_{ur,j}^{~~2}$ (TAM). The gravitational wave equation is then sourced by the shear
$\pi_{ur} = \frac{8}{5} \Theta_2^2$, replaced by eq.~(B.27) of TL13 in the optimal case.

For scalar modes, we implemented the TAM hierarchy in both the synchronous and newtonian gauge. In the next section, we show comparison plots obtained in the synchronous gauge, but we checked explicitly that the curves are identical in the newtonian gauge.

\subsection{Accuracy of the hierarchies}

We turn to the evaluation of the difference between both hierarchies
in the curved case. Since the optimal hierarchy is mathematically valid only in
the flat case, we expect differences proportional to $|\Omega_K|$ in the angular
spectra. In principle, some cancellations could occur such that the difference would scale with a higher power of $\Omega_K$; but we checked explicitly that this is not the case: the differences between the CMB spectra computed by CLASS in the two schemes scale indeed linearly with the curvature density fraction.

\begin{figure}[!htb]
\includegraphics[width=\columnwidth]{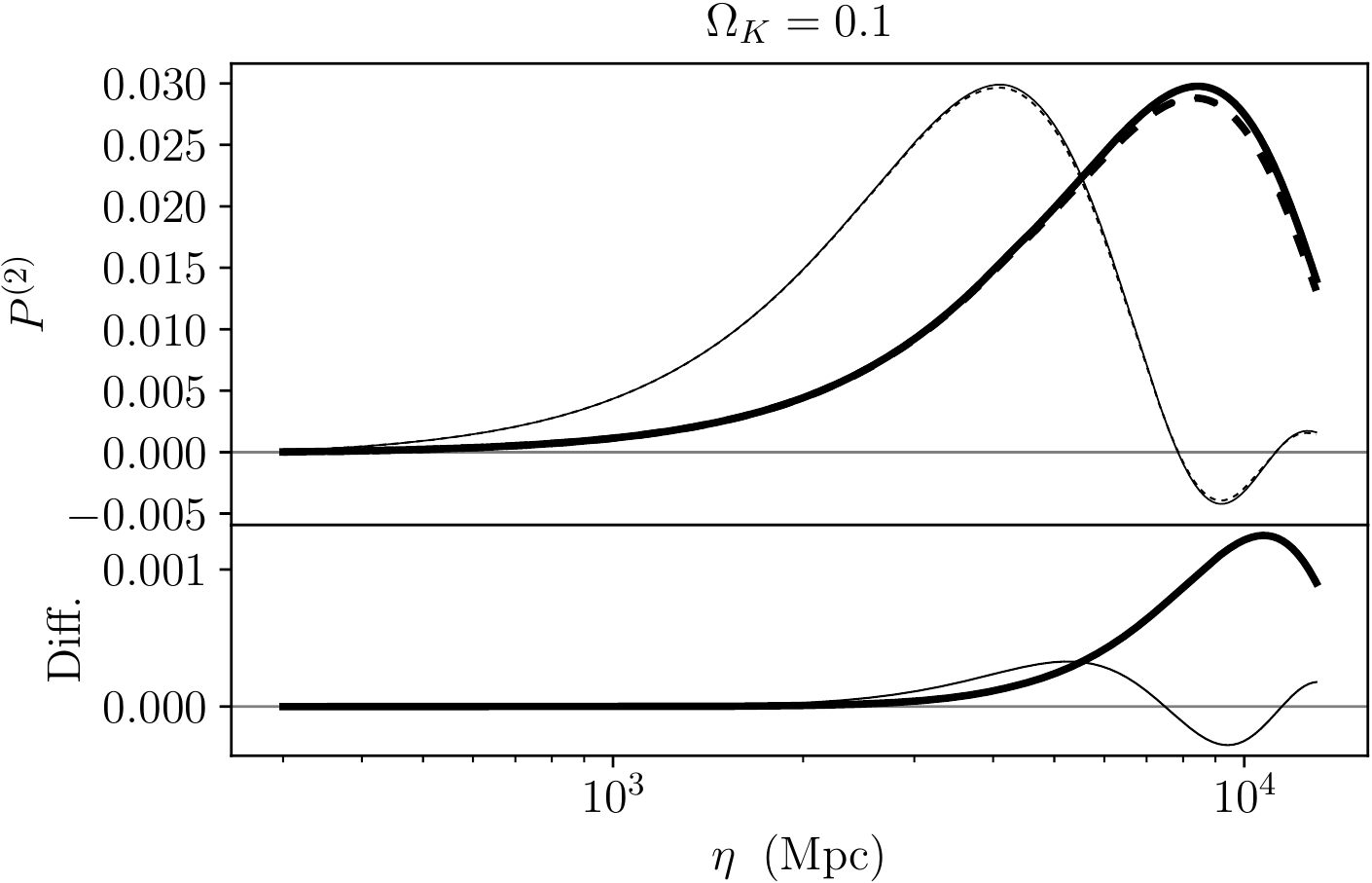}
\caption{Sources for tensor modes $P^{(2)}$ used in the line of sight
  method. The continuous line are computed with the optimal hierarchy, and
  the dashed lines with the the TAM hierarchy. The thicker lines are for the
  mode $k=0.0005\,{\rm Mpc}^{-1}$, and the thiner lines are for
  $k=0.001\,{\rm Mpc}^{-1}$. We only show the case of negative
  curvature with $\Omega_K=0.1$, but positive curvature sources are extremely similar. The lower
  panel shows the difference between the curves of the upper panel.
}\label{Plot1a1b}
\end{figure}

Both implementations rely on the line-of-sight integral with identical radial functions. Since these functions account for projection effects from $q$-space to harmonic space, the geometrical effects induced by curvature -- that govern, for instance, the angular scale of the acoustic peaks -- are correctly accounted for in the two approaches. Differences can only arise from slightly different values of the source functions that appear in eqs.~\eqref{IntSol}: $^\Theta{\cal C}_j^m(\gr{q},\eta)$, ${\cal G}_j^m(\gr{q},\eta)$ (with $j=0,1,2$) and
$^E{\cal C}_2^m(\gr{q},\eta)$ in the two schemes. Figure \ref{Plot1a1b} shows such differences at the level of the tensor polarization source function $P^{(2)}$ of eq.~\eqref{CollQiU}, which is related to the sources of eqs.~\eqref{IntSol} through
$^E{\cal C}_2^2=-\sqrt{6}\tau' P^{(2)}$.

In each of the two schemes, the source functions are derived from equations that are sensitive to curvature only through:
\begin{enumerate}
\item coefficients ${}_s \kappa_j^m$ in the hierarchies, that involve factors like $\sqrt{1-nK/q^2}$ for various integers $n$, 
\item initial conditions,
\item the background evolution at very small redshift. 
\end{enumerate}

Since the two schemes share the same initial conditions and background evolution, and since they have a common flat-space limit, differences can only be caused by $\sqrt{1-nK/q^2}$ factors. Thus these differences must be more significant at large wavelengths, that is, for small multipoles.

\begin{figure*}[!htb]
\includegraphics[width=0.49\linewidth]{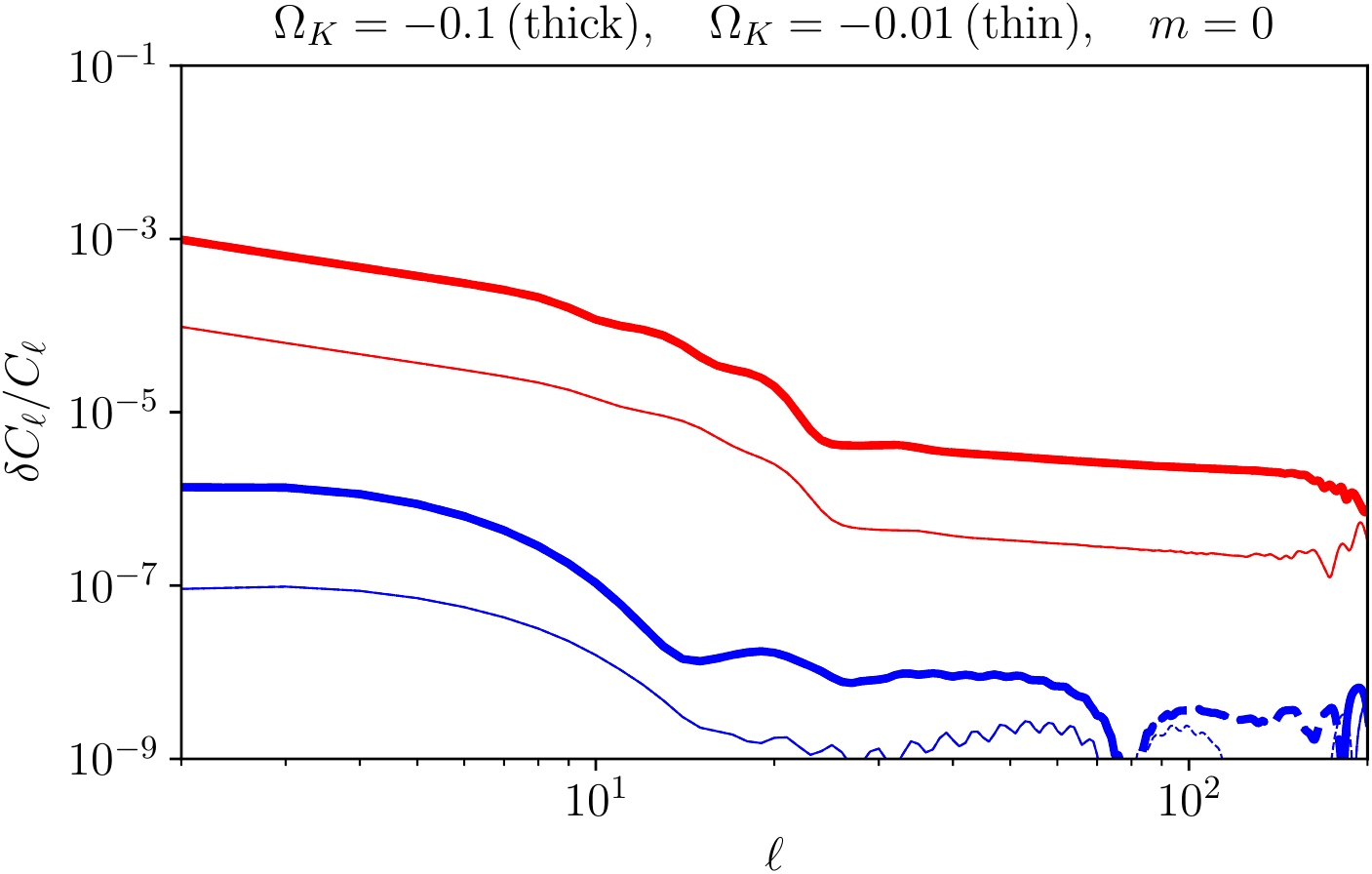}
\includegraphics[width=0.49\linewidth]{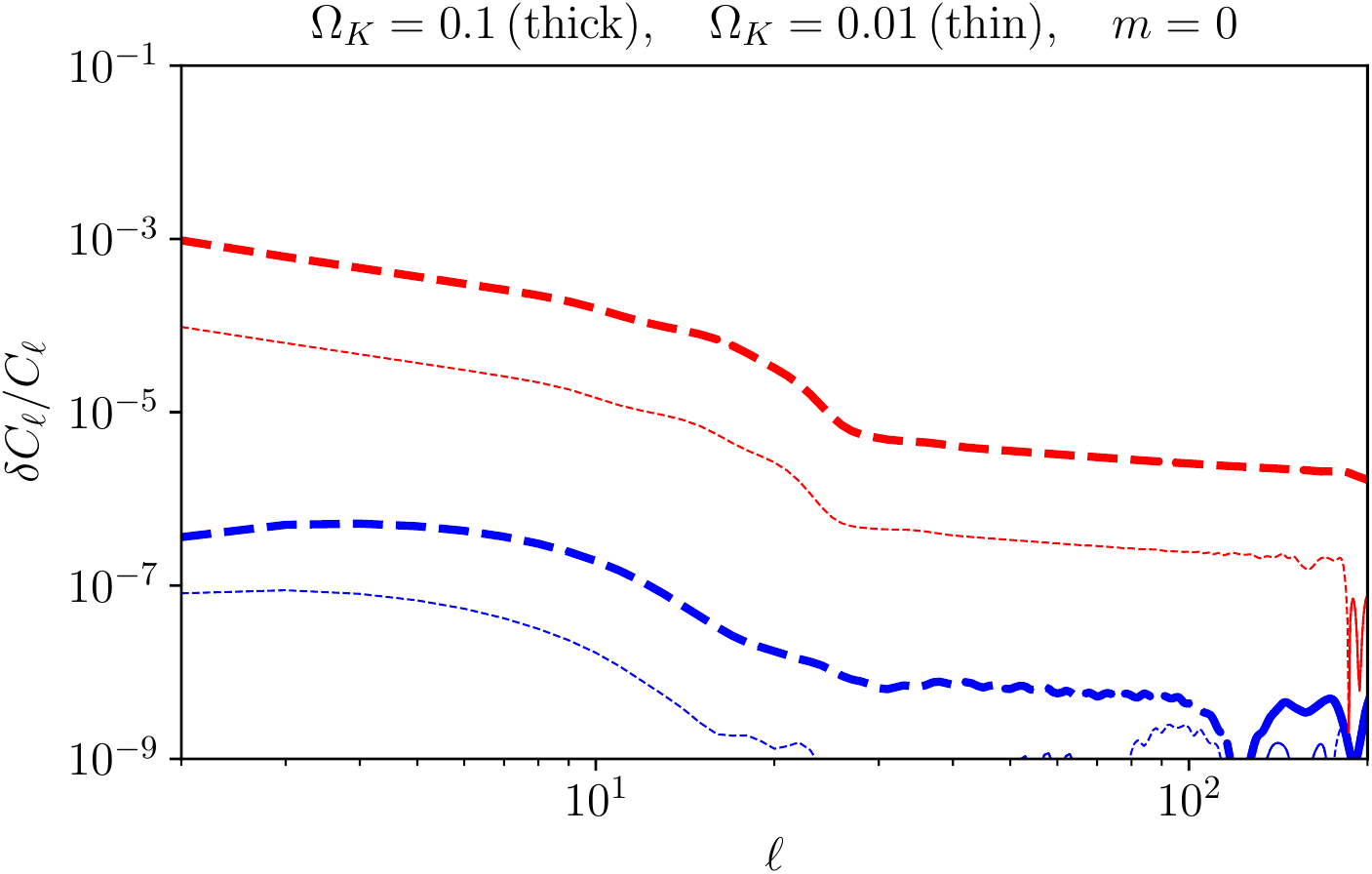}\\
\includegraphics[width=0.49\linewidth]{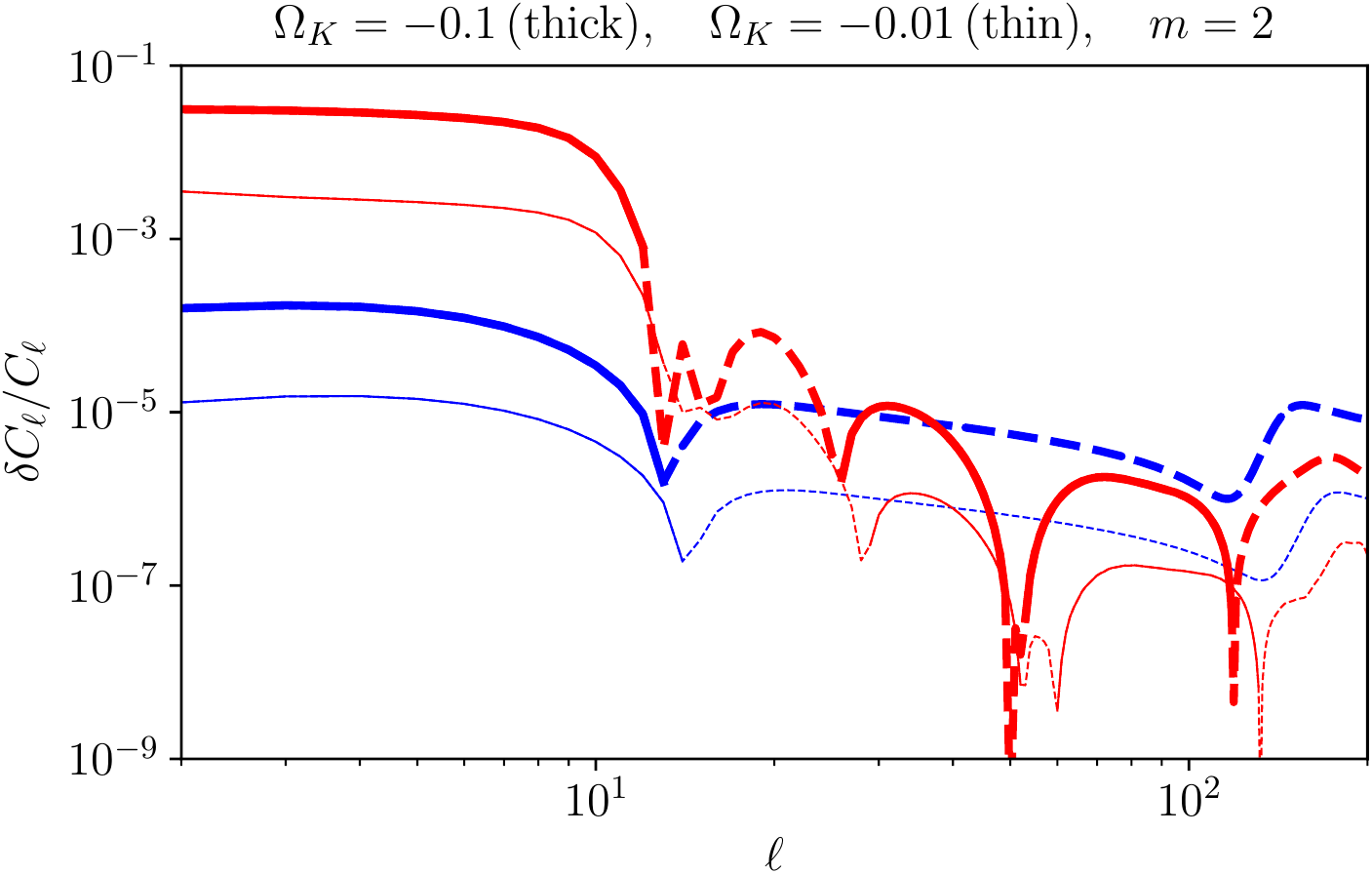}
\includegraphics[width=0.49\linewidth]{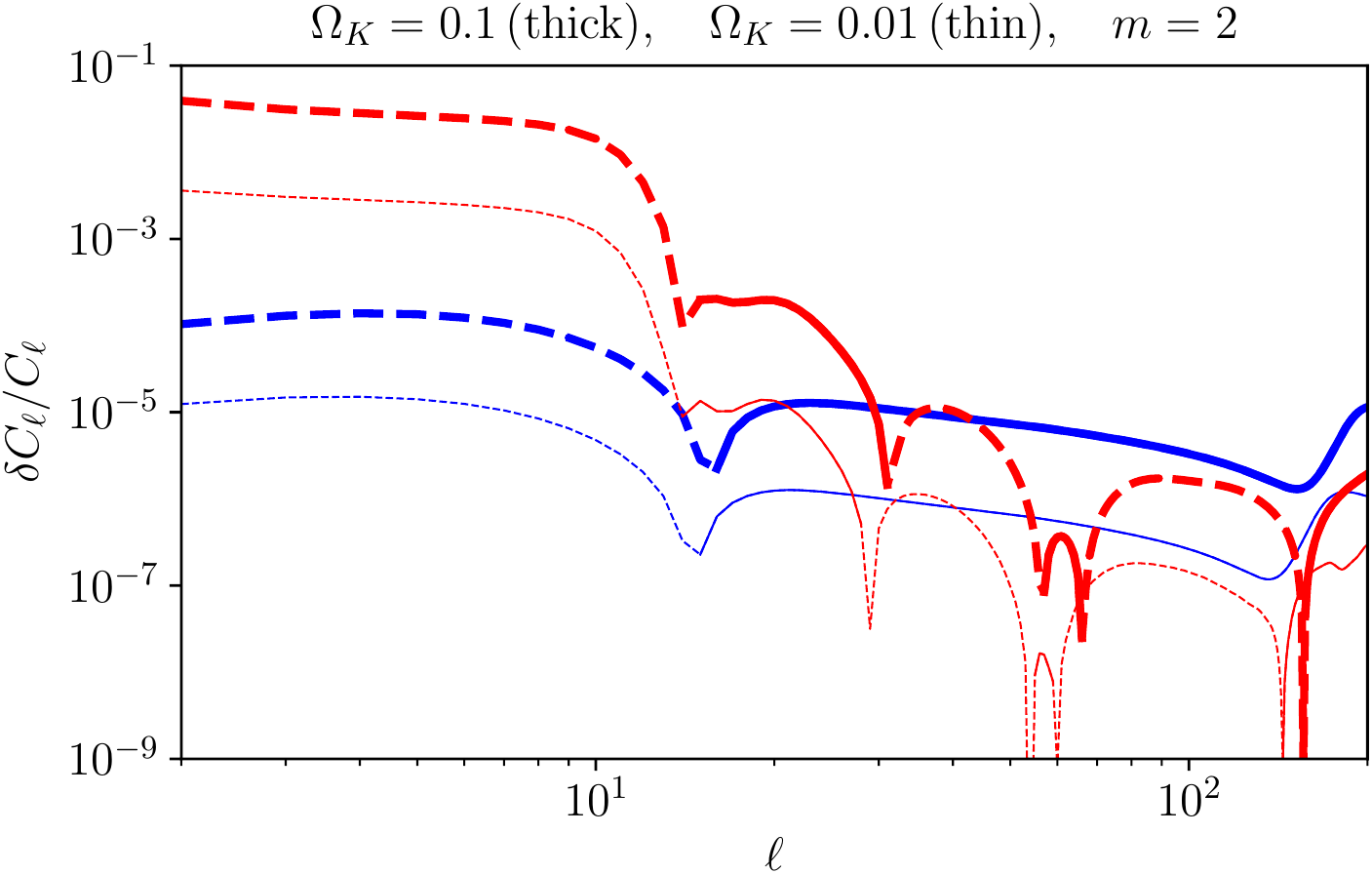}
\caption{Relative differences of spectra $C_\ell^{\rm
    optimal}/C_\ell^{\rm TAM}-1$ for $TT$ spectra [blue (lower) lines] and $EE$ spectra [red (upper)
  lines]. Left panels are with positive curvature
  $\Omega_K<0$, and the right panels with negative curvature
  $\Omega_K>0$. Positive values are in continuous lines, and negative
  values in dashed lines. Top and bottom panels are for scalar and tensor perturbations, respectively.  The cosmological parameters are the one of the
last column in table 2 of \cite{Aghanim:2018eyx}, except for
 the modification in $\Omega_K$ which is accompanied by a modification in $\Omega_{\Lambda}$.}\label{Plot2a2b2c2d}
\end{figure*}

The terms related to photon multipoles in the source functions of eqs.~\eqref{IntSol} are all multiplied by the visibility function $-\tau' e^{-\tau}$, which peaks around the times of recombination and reionization. We expect the differences between the hierarchies to manifest themselves more clearly around the time of reionization. Indeed, on the last scattering surface, the sources emerge from the tight-coupling regime, while at reionization free-streaming has entirely shaped them. Since the major difference between the hierarchies is the treatment of free-streaming, we expect that they have more impact on contributions from reionization. However, this contribution is subdominant in angular spectra, excepted for polarization spectra at low $\ell$. This induces a global suppression of differences, excepted on scales corresponding to the reionization bump in the polarization spectra. 

Furthermore, there are several properties which conspire to eventually reduce even more the differences in angular spectra (see Figs.~\ref{Plot2a2b2c2d}) which we now detail.

For scalar temperature, we have already seen that the Boltzmann hierarchy of the two schemes are equivalent (because all quantities are expanded along the same normal modes $_0M_j^0$), up to the term that couples the temperature and polarization hierarchies. This term is part of $^\Theta{\cal C}_j^0$ in eq.~\eqref{Hierarchy}, and proportional to $P^{(0)}$. Intuitively, it represents the flow of power from temperature to polarization induced by Thomson scattering. The CMB is known to be only slightly polarized, precisely because this flow is very small.
Since the different polarization hierarchies only affect the temperature hierarchy through this term, the difference they induce on the evolution of temperature multipoles is very small. 
Finally, the scalar temperature spectrum $C_\ell^{TT}$ is inferred from the scalar temperature source function of eq.~\eqref{IntSol}, which depends mainly on temperature multipoles, on the baryon velocity field and on metric perturbations; the electric quadrupole moment $E_2^0$ brings only a very small correction. Thus we expect a very minor impact of polarization errors on the scalar temperature spectrum.

This is confirmed by the blue (lower) curves in the top panels of Figure~\ref{Plot2a2b2c2d}. The difference between the scalar spectra $C_\ell^{TT}$ predicted by the two hierarchies peaks at small $\ell$'s, and is at most of the order of $10^{-5}|\Omega_K|$. We checked explicitly that most of the difference comes from the value of the source terms (and in particular of the quadrupole $E_2^0$) around the time of reionization: in a cosmological model with reionization switched off, the difference between the spectra is orders of magnitude smaller.

For scalar polarization, we do expect a larger difference, because the Boltzmann hierarchies of the two schemes are not anymore exactly equivalent for $\Omega_K \neq 0$. For $m=0$, in the TAM scheme, $B_j^0$ multipoles are not sourced and remain null. Equation (B.11) of TL13 gives an explicit relation between $E_j^0$ and $G_j^0$, but according to our previous discussion, this relation would be exact only for separable normal mode functions ${}_sM_j^0$, that is for $K=0$. The source function in the polarization line-of-sight integrals, $^E{\cal C}_2^0=-\sqrt{6}\tau' P^{(0)}$, involves the sum 
\be
P^{(0)}=(\Theta_2^0 -\sqrt{6} E_2^{0})/10~.
\ee
In flat space, using (B.11) of TL13, $\sqrt{6}  E_2^0$ would be exactly equal to $-\frac{5}{4} (G_0^2+G_2^2)$. In curved space, one can explicitly check that the term $\sqrt{6}  E_2^0$  coming from the solution of the $E_j^0$ hiearchy and the term $-\frac{5}{4} (G_0^2+G_2^2)$ coming from the solution of the $G_j^2$ hierarchy differ by $\sqrt{1-nK/q^2}$--like factors.
However, in both schemes, $P^{(0)}$ is dominated by the contribution of the temperature quadrupole, correctly given in the two schemes by $\Theta_2^0=5/4F_2^0$, and to which the polarization multipoles only bring a small correction. Since the temperature hierarchy is almost unaffected by errors in the optimal scheme, differences in the solution of the polarization hierarchies do not fully propagate to the polarization spectra. This explains why the error on $C_\ell^{EE}$, shown in the red (upper) curves in the top panels of Figure~\ref{Plot2a2b2c2d}, is still very small, of the order of $10^{-2}|\Omega_K|$ at small $\ell$'s. It is however~$\sim 10^3$ times larger than the largest difference for the temperature. Since the difference between the two schemes has more impact at the reionization epoch, the residuals are the largest in the range $\ell \leq 20$ corresponding to the reionization bump in $C_\ell^{EE}$.

For tensor modes, differences are expected to be even larger, since in that case, both temperature and polarization hierarchies are different. However the impact of the hierarchies is reduced again by another consideration in the temperature case. The tensor temperature source function in eqs.~\eqref{IntSol} is given by the sum $-H'+\kappa' P^{(2)}$, where $H$ is the gravitational wave transfer function. As already seen in Figure~\ref{Plot1a1b}, the term $P^{(2)}$ is clearly sensitive to the difference between the hierarchies, especially around the time of reionization. However, the tensor temperature power spectrum is dominated by the term $-H'$ that represents an integrated Sachs-Wolfe effect  caused by gravitational waves. This term is given by the same Einstein equations in the two schemes, and only depends very weakly on the choice of hierarchy, in spite of the small back-reaction of photon and neutrino shear on $H$. Thus, once more, we find a very small impact of the optimal hierarchies on the tensor temperature spectrum, of the order of $10^{-3}|\Omega_K|$ (see the blue (lower) curves in the bottom panels of Figure~\ref{Plot2a2b2c2d}). 

Finally, for tensor polarization, the source term in eqs.~\eqref{IntSol}  is only given by $P^{(2)}$, and thus by temperature and polarization mutipoles. This is the only case in which we find that the optimal hierarchy induces a potentially relevant error, of the order of $0.5|\Omega_K|$ for $C_\ell^{EE}$ and $\ell \leq 10$ (see the red (upper) curves in the bottom panels of Figure~\ref{Plot2a2b2c2d}). 
The range $\ell \leq 10$ coincides with the reionization bump in the tensor $C_\ell^{EE}$, which is again consistent with the fact that the difference between hierarchies has more impact around reionization than recombination.
We find essentially identical results for the $C_\ell^{BB}$ spectrum. 

Since in these sections we were interested in extremely small differences between the angular spectra, we ran CLASS with enhanced accuracy settings (namely, the ones of the public precision parameter file {\tt cl\_ref.pre}). Note that even with such settings, a comparison with the CAMB code suggests that both Einstein-Boltzmann solvers are accurate at least at the $10^{-4}$ level~\cite{Lesgourgues:2011rg}. However, the level of convergence of each of the two codes against an increase in their own precision parameters is much better than that. Thus showing residuals smaller than $10^{-4}$ is still meaningful when one wants to highlight the effect of just one type of error (in our case, the one induced by the optimal hierarchy). Even when the residuals shown in Figure~\ref{Plot2a2b2c2d} are below $10^{-4}$, they show the specific impact of switching between hierarchies, even in the presence of comparable or larger sources of errors in other aspects of the code. To check this, we tried several accuracy settings between default precision and {\tt cl\_ref.pre}, and found that our residuals are stable and well-converged at least for $\ell<200$. For $\ell>200$ this was not always the case and we choose to limit the plots to the first range. But given 
that there is a solid analytical argument for the error to decrease with $\gr{q}$ and $\ell$, it is sufficient to obtain converged results for small multipoles.

For scalar modes, we plotted the results obtained using the synchronous gauge, but we found nearly identical curves when running CLASS in the newtonian gauge. The residuals are nearly equal even in ranges when the error induced by the optimal hierarchy is smaller than the one induced by the newtonian gauge (the two gauges agree at the level of $10^{-6}$). This brings further confirmation that our residuals correctly capture the error induced by the optimal hierarchy only.

\subsection{Efficiency of the hierarchies}

To compare the efficiency of the two approaches as implemented in CLASS, we need to make several choices. Indeed, the result of timing tests should 
depend on many factors like:
\begin{itemize}
\item the level of precision: high precision (in particular, a larger truncation multipole $j_{\rm max}$) is more favorable to the optimal hierarchy; the choice of algorithm to solve Ordinary Differential Equations (ODEs) is also important;
\item the underlying cosmology: with more ingredients involved, the weight of photons in the system of perturbation equations decreases, and the difference between hierarchies is less pronounced;
\item the timing method: if we compare the total execution time of the code in the two cases, the result will depend a lot on the requested output; this is not the case if we only compare the time $\Delta t_{\rm ODE}$ spent by CLASS in the  loop over $q$-modes, during which the system of ODEs is integrated over time for either scalar or tensor perturbations;
\item the chosen number of parallel threads, the compiler, the optimization flags, etc.
\end{itemize}
Here we will focus on the ratio of $\Delta t_{\rm ODE}$ when the two hierarchies are used, while running CLASS with default precision, and thus with the {\tt ndf15} ODE solver \cite{ClassII}. The default precision settings of the current version of CLASS have been optimized for accurate MCMC fits of Planck data. For the optimal hierarchy, they give $j_{\rm max}=12$ for scalar temperature, 10 for scalar polarization and 5 for tensor temperature and polarization. In the TAM hierarchy, to be consistent, we should keep the same truncation for scalar modes and increase $j_{\rm max}$ by two for tensor modes\footnote{
Indeed, the functions
 $Y_m^m$ and $\tilde{\epsilon}^m$ that appear in the relations between the expansions in equations  (\ref{DefineFjmHierarchy}, \ref{DefineGjmHierarchy}) have the geometry of a monopole for scalar modes and of a quadrupole for tensor modes.}. Thus we set $j_{\rm max}=7$ by default in the TAM tensor case. By comparing with the results of the previous sections (obtained with high precision settings and $j_{\rm max}=50$ in all cases), we checked that with such default precision settings, the accuracy level is roughly the same in the two schemes.

In these timing tests, we assumed a $\Lambda$CDM model with massless neutrinos, spatial curvature and tensor modes, and we asked only for CMB output (in CLASS syntax, {\tt output = tCl,pCl,lCl}). Our results are however independent of $\Omega_K$ and apply also to flat models. We quote relative differences when the code is run sequentially, using the compiler {\tt gcc 9.2.0} with option {\tt -O4}.

We find that for scalar modes, the time interval $\Delta t_{\rm ODE}$ is the same in the two schemes up to negligible (percent level) differences. This is consistent with the fact that the two hierarchies involve roughly the same number of photon mutipoles: 13+11=24 in the optimal case, and 13+9=22 in the TAM case since the scalar magnetic mutipoles $B_j^0$ vanish and do not need to be defined.
For tensor modes, we find a 13\% speed up in the optimal scheme. In that case, the optimal hierarchy involves 12 multipoles and the TAM hierarchy 18 multipoles.

Thus, with a line-of-sight method and standard precision requirements, the efficiency of the two schemes is very similar. Choosing one of them is mainly a matter of taste. Given that the optimal hierarchy is accurate enough for most purposes, in our implementation, we kept it as the default choice for continuity with previous CLASS versions.

\section*{Conclusion}

The incorrect relation between the Stokes parameters $Q$ and $U$ assumed by the optimal hierarchy leads to different source terms in the line-of-sight integrals, especially around the time of reionization, when the sources are shaped by the details of the free-streaming solution. In the observable angular spectra, differences remain very small, because the reionisation epoch accounts only for a small part of the total spectra. They are further suppressed for scalar modes by the dominant role of temperature multipoles, correctly handled by both hierarchies, and for tensor temperature by the dominant role of metric perturbations. They are thus predominately seen in the tensor polarization spectra, on the scale of the reionization bump ($\ell \leq 10$).

For instance, for $|\Omega_K|=0.1$, the tensor polarization spectra
are affected at the $5\%$ level for such multipoles. In the future, if
cosmological observations came to prefer a slightly curved universe
with, for instance, $|\Omega_K| \sim 0.02$, using the TAM hierarchy
instead of the optimal one would be important for reconstructing the
tensor-to-scalar ratio from $C_\ell^{BB}$ with an accuracy of
1\%. However, if the current bound from Planck+BAO gets confirmed,
$|\Omega_K|<0.002$ (68\%CL), the optimal hierarchy is sufficient to
guarantee a 0.1\% accuracy on the tensor polarization spectra. On the
other hand, if one is interested  on the transfer of super-Hubble or
supercurvature modes,  then it is crucial to rely on the TAM hierarchy. For instance, it has been shown that very
long (i.e., maximal wavelength) modes on the top of isotropic spacetimes are equivalent to Bianchi universes~\cite{Pontzen2010,Duality}, and in
that case it is crucial to rely on the correct TAM hierarchy to infer
the observational consequences with this
approach~\cite{PPLetter}. Using the TAM hierarchy might also be of
importance for checking the validity of consistency theorems in single field inflation~\cite{Maldacena:2002vr}, which allows one to connect the primordial bispectra in squeezed configurations to products of the primordial spectra \cite{Creminelli:2011sq,Mirbabayi:2014hda}, since it involves very large-scale modes modulating the small scales dynamics.

\acknowledgements{We would like to thank Thomas Tram and Nils Sch\"oneberg for very valuable help and comments. TP thanks the Brazilian Funding Agency CNPq (grants 311527/2018-3 and 438689/2018-6) for the financial support.}

\bibliography{CLASS-Comparison}

\end{document}